\documentstyle[aps,preprint]{revtex}
\input{psfig.sty}
\renewcommand{\today}{September 18, 1997}
\begin{document}
\title{Surface Transitions for Confined Associating Mixtures} 
\author{Michael Kotelyanskii,\cite{emailmike} Sanat K. Kumar
\cite{emailkumar}$^{,\star}$}
\address{Department of Materials Science and Engineering, 
Pennsylvania State University,  
University Park, PA 16802.}

\date{\today}
\maketitle

\begin{abstract}
Thin films of binary mixtures that interact 
through isotropic forces and directionally
specific ``hydrogen bonding" are considered through
Monte Carlo simulations. We show, in good agreement with experiment,
that the single phase of these mixtures can be
stabilized or destabilized on confinement.
These results resolve a long standing controversy,  
since previous
theories suggest that confinement only
stabilizes the single phase of fluid mixtures.
\end{abstract}

\newpage
The phase behavior of mixtures in thin films can be significantly
different 
from the bulk 
\cite{Fish0,Fish1,Fish2,NakanishiFisher,Fish3,Mader76,Meadows,Scheibner,PSPVME,russians,BinderReview,Lubensky}.
Fisher et al. theoretically considered 
an incompressible
fluid mixture which interacted with
isotropic nearest neighbor forces 
and showed that  
confinement only stabilized the single phase 
\cite{Fish1,Fish2,NakanishiFisher,Fish3}. 
This prediction is robust, and cannot be altered by the magnitude of
the interaction between the fluids and the walls.

Experiments, however,  
show that,
depending on the surfaces, 
confinement can either stabilize or destabilize the single phase relative
to the bulk  
\cite{Meadows,Scheibner,PSPVME,russians}. 
Since the model of Fisher corresponds to the simplest case of
a binary fluid with isotropic nearest
neighbor forces, 
a model with more complex
interactions could rationalize the experiments
\cite{Fish3}.
We have performed Monte Carlo simulations on mixtures which interact with
isotropic nearest neighbor interactions and directionally specific 
bonds. 
The competition between the unfavorable dispersive interactions,
and the favorable ``hydrogen bonds" (HB) leads to closed loop
phase diagrams in the bulk. While the bulk behavior of HB systems
has been well explored
\cite{Hirshfelder,Barker52,wheeler,walker,Boris89},
their behavior in confined
geometries has not been studied theoretically.
We find that, when these systems are confined between
plates which only interact through HB
with the molecules, both stabilization [``ordinary"] and
destabilization [``surface transition"] behavior can occur 
depending
on the specific parameters employed.
To our knowledge this represents the 
first theoretical evidence for surface transitions 
for a confined binary mixture. 

A binary fluid mixture
with isotropic nearest neighbor interactions is
isomorphic with the
spin 1/2 Ising model. 
Consider an Ising system confined between
two symmetric walls which are $D$ layers apart.
$J$ and $J_{1}$ are the spin coupling parameters in the bulk
and in  the surface layers, respectively
\cite{Fish0}. 
$\Delta \equiv \frac{J_{1}-J}{J}$ has a special value, 
$\Delta_{c}$, so that when $\Delta < \Delta_{c}$
the phase mixed state is stabilized in the thin film (``ordinary"
behavior) \cite{Fish0,BinderReview}. In contrast,
when $\Delta > \Delta_{c}$ the surface undergoes critical ordering
even when the bulk is phase mixed (``surface transition") \cite{desai,kardar}. 
For the case of simple mixtures
it has been shown that $\Delta < \Delta_{c}$ and hence only ordinary
behavior is predicted \cite{Fish0,Fish1,Fish2,NakanishiFisher,Fish3}. 
We shall show here that the introduction
of a new interaction, the HB, could, under certain circumstances
result in $\Delta > \Delta_{c}$, and thus
yield complex surface behavior
that is consistent with experiment.

We consider a completely filled cubic lattice with its sites  
occupied by either $A$ or $B$ molecules. 
We model {\em symmetric} mixtures where the interactions
between any two $A$ molecules, and any two $B$ molecules are identical.
The molecules
interact through isotropic nearest neighbor interactions 
characterized by the energy scale: $\chi \equiv
(d/2k_{B}T) (2\epsilon_{AB}-\epsilon_{AA}-\epsilon_{BB})$. 
$d$ is the coordination number (=6), and $\epsilon_{ij}$
is the interaction energy between a nearest neighbor $i-j$ pair.
Therefore, $\chi=6/T^{\star}$ [$T^{\star}=k_{B}T/
(\epsilon_{AB}-\epsilon_{AA})$].
Each molecule has 
one ``donor" and one ``acceptor" which can participate in
nearest neighbor HB.  
Since the molecules are structureless 
the donors and acceptors do not have prespecified
locations.
The HB interactions are described by two
equilibrium constants $k_{AA}$ and $k_{AB}$ 
for the bonds between an $A$ donor and an $A$ acceptor on 
different molecules, 
and for bonds between $A$ and $B$
particles [either one being the donor], respectively. 
$k_{ij} \equiv P_{ij} e^{-E_{ij}/k_{B}T}$ \cite{entropy}.
We employed $k_{AA}=0.0275e^{1.8/T^{\star}}$, 
and $k_{AB}=0.0134e^{4.5/T^{\star}}$. 
While we have explored the phase behavior for a range of values  of
$\chi$, $k_{AA}$ and $k_{AB}$, we focus on this one 
set of parameters which
yields closed loop phase behavior for the
bulk mixture.

In simulations of the bulk
\cite{MKBulk} we employed periodic boundary conditions 
in all three directions.  
For thin films the periodic boundary conditions  
along the $z$ direction were replaced by two symmetric hard walls. 
The walls could only interact with the molecules through HB, i.e.,  
$k_w \equiv P_{w}e^{-E_w/k_{B}T}$ independent of the identity of the molecule.
The Monte Carlo simulations utilize the symmetry of the system
and locate phase coexistence through the semigrand 
ensemble method \cite{SemiGrand} with the condition that
$\Delta \mu \equiv \mu_{A} - \mu_{B} =0$. $\mu_{i}$ is the chemical
potential of species $i$. An elementary
Monte Carlo move is to 
change the identity of a randomly chosen particle. The move
is accepted following the standard Metropolis criterion \cite{SemiGrand}. 
Another elementary move consists of the creation or elimination of a 
HB.        
The composition of the mixture [$x=N_A/(N_{A}+N_{B})$, where 
$N_{i}$ is the number of $i$ particles]
is variable and the binodal can be determined from 
histograms of its distribution, $P(x)$. 
When the system is miscible 
$P(x)$ has a maximum at $x=1/2$. 
In the immiscible regime  
two maxima are observed at $x=1/2 \pm x_{b}$, 
corresponding to the coexisting phases. 

\begin{figure}
\centerline{\psfig{figure=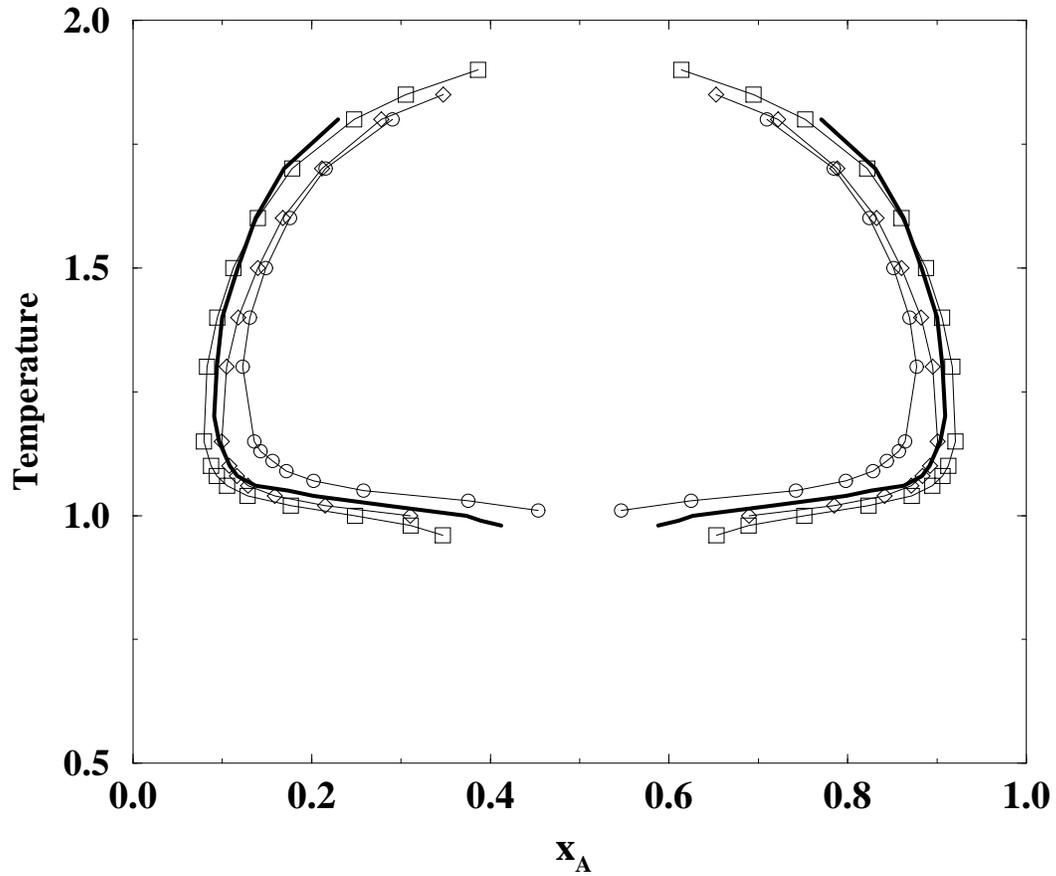,width=16cm,angle=-90}}

\vspace{0.2in}
\caption{Binodals for the bulk system (thick solid line), and for the $D=16$ film with 
$k_w=0.0134e^{10/T^{\star}}$ (squares),  $k_w=0.0134e^{4.5/T^{\star}}$ (diamonds), and
without specific interactions with the wall, (circles). }
\label{BulkBinodal}
\end{figure}

In Figure \ref{BulkBinodal} the binodals for the bulk and
for three films of $D$=16  
are shown. The bulk system displays closed
loop phase behavior, a feature that is characteristic 
of many HB mixtures 
\cite{Hirshfelder,Barker52,wheeler,walker,Boris89}. 
If one defines the critical temperatures as the maxima of susceptibility 
\cite{BinderHeermann}, then, the upper critical solution 
temperature [UCST] is at $T^{\star}_{ucst}=1.94\pm0.04$,
while the lower critical solution temperature [LCST] is
at $T^{\star}_{lcst}=1.02 \pm 0.02$ [see Figure 3]. 
This result is obtained for system sizes of 8$\times$8$\times$8 and 
16$\times$16$\times$16, and appears
to be effectively independent of system size in this range.
Note that a system with no HB interactions
reduces to a standard 3-d Ising model with a critical
temperature of $T_{0}^{\star}=2.25$.

We now consider
this HB mixture when it is confined between two hard, non-interacting
walls.
The mixed state is stabilized and the film binodal
lies ``inside" the bulk binodal.
$T^{\star}_{ucst}=1.80\pm0.05$
and $T^{\star}_{lcst}=1.03\pm0.02$.
This behavior is in line with previous predictions that simple 
systems only show ``ordinary" behavior when confined
\cite{Fish0,Fish1,Fish2,NakanishiFisher,Fish3,BinderReview}.
When specific interactions are allowed between the molecules and 
the walls, the phase behavior of the film 
changes {\em qualitatively}. This is the essential point of our paper. 
The two other curves in Figure \ref{BulkBinodal} correspond to 
$k_w= 0.0134e^{4.5/T^{\star}}$ and  
$k_w= 0.0134e^{10/T^{\star}}$, respectively.  
In both  cases the LCST is destabilized, and the UCST is
stabilized on confinement. 
In the first case $T^{\star}_{ucst}=
1.90\pm 0.05$,
$T^{\star}_{ucst}=0.98\pm 0.02$, while
$T^{\star}_{ucst}=1.90\pm 0.05$, $T^{\star}_{lcst}=0.99\pm 0.02$
in the second case.

\begin{figure}
\centerline{\psfig{figure=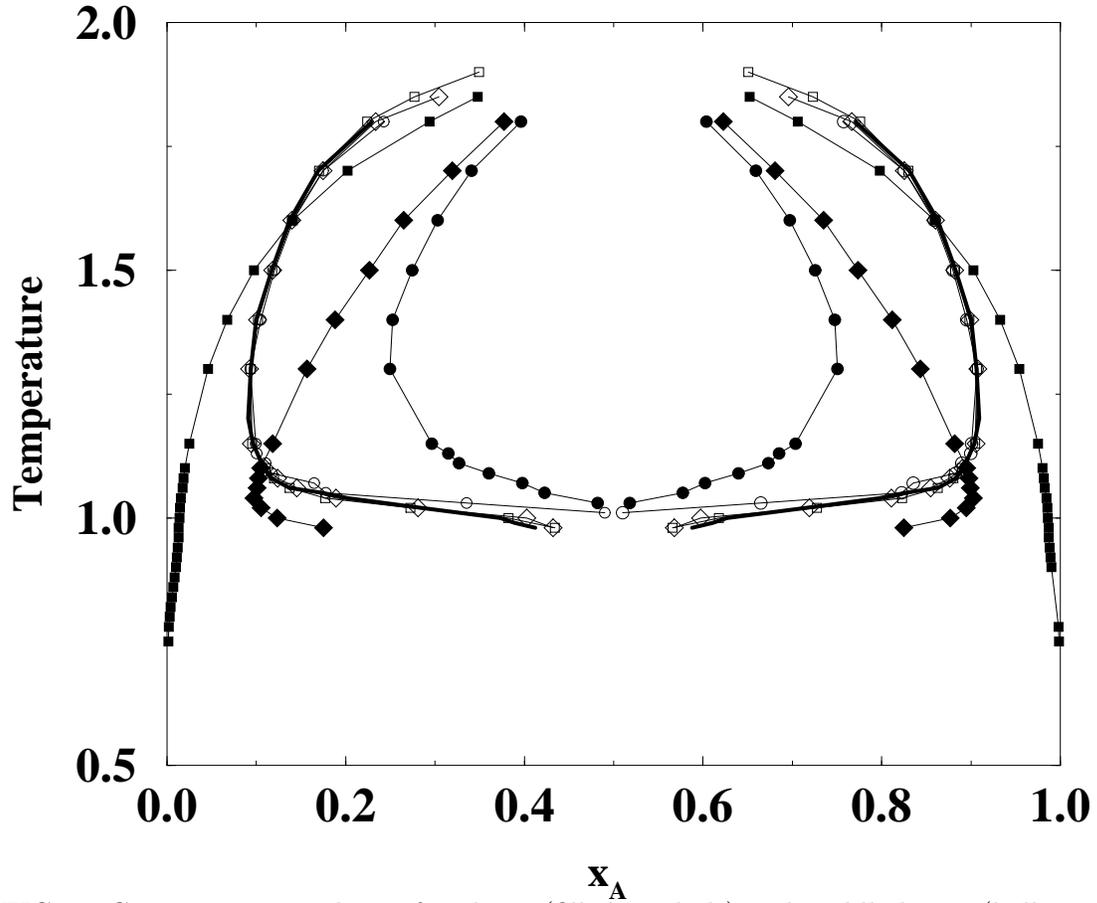,width=16cm,angle=-90}}

\vspace{0.2in}
\caption{Compositions in the surface layer (filled symbols) and middle layers (hollow symbols) 
of the $D=16$ film at the same conditions as fig. \ref{BulkBinodal}. 
Circles - neutral walls with no interactions, squares and diamonds - walls with specific interactions. 
Squares -- $k_w=0.0134e^{10/T^{\star}}$, and diamonds -- $k_w=0.0134e^{4.5/T^{\star}}$.
Bulk binodal (thick solid line, same as fig. \ref{BulkBinodal}) is shown for the reference.} 
\label{FigSurface}
\end{figure}

\begin{figure}
\centerline{\psfig{figure=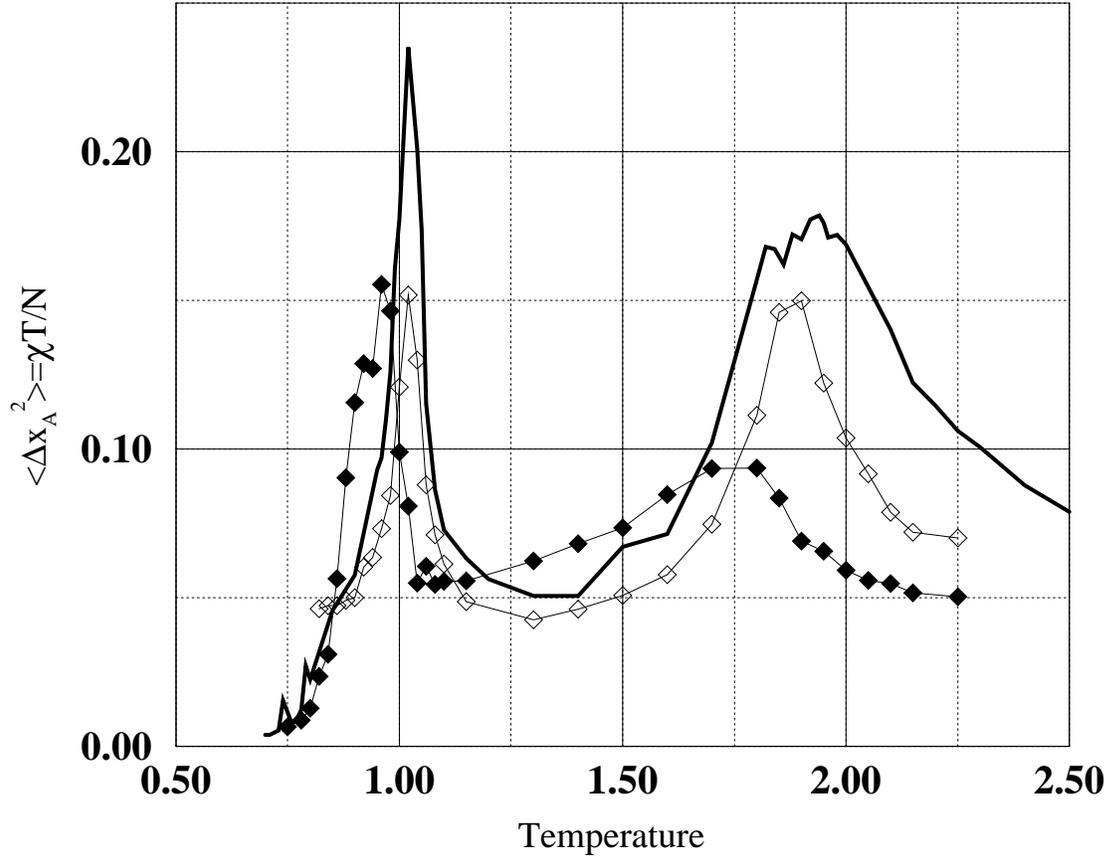,width=16cm,angle=-90}}

\vspace{0.2in}
\caption{Susceptibilities $\chi$, obtained from the mean squared deviations
of $x_A$, vs. temperature for $D=16$ film 
with $k_w=0.0134e^{4.5/T^{\star}}$. Filled diamonds
-- surface layers ($z=0$, and $z=15$),
Hollow symbols --
middle of the film ($z=7$, and $z=8$).
Susceptibility with no surfaces and periodic boundary conditions is 
plotted for the reference (thick solid line).
}
\label{FigSuscept}
\end{figure}

To understand 
these issues better, in Figure \ref{FigSurface}
the composition in the surface layer, 
and in the center of the $D=16$ films are plotted.
Only composition
values different from $x=1/2$ are plotted for clarity.
In all cases the data from the center of the film
virtually coincide with the bulk binodal.
When the walls are neutral, 
the composition in the surface layer is closer to 0.5 than in the bulk. 
This is consistent with the ``ordinary" transition behavior observed
in this case. 
Similar behavior is observed near the UCST with the interacting walls.
In contrast, in the vicinity of the LCST the 
surface is ``more ordered" than the bulk, a signature of
``surface transition" behavior.
Figure \ref{FigSuscept} shows plots of
susceptibility as a function of temperature 
for the middle and surface layers in the $D=16$ 
film with $k_w=0.0134e^{(4.5/T^{\star})}$.
The susceptibility of the middle layer tracks 
the bulk behavior, consistent with trends
observed in Figure 2. 
In the vicinity of the LCST the
surface layers show a distinct peak, consistent with the
notion of a  ``surface transition".
Notice also  that in the case with strongest wall interactions, 
the film surface remains 
ordered even when the middle layers becomes mixed below the LCST.

The transition from ``ordinary" behavior around the UCST to the 
``surface transition" at the LCST occurs through   
the ``extraordinary transition", when the surface composition equals
that in the middle of the film.
This occurs at $T^{\star}\approx1.10$ 
for $k_w=0.0134e^{(4.5/T^{\star})}$, and
at $T^{\star}\approx1.60$ with $k_w=0.0134e^{(10.0/T^{\star})}$. 

\begin{figure}
\centerline{\psfig{figure=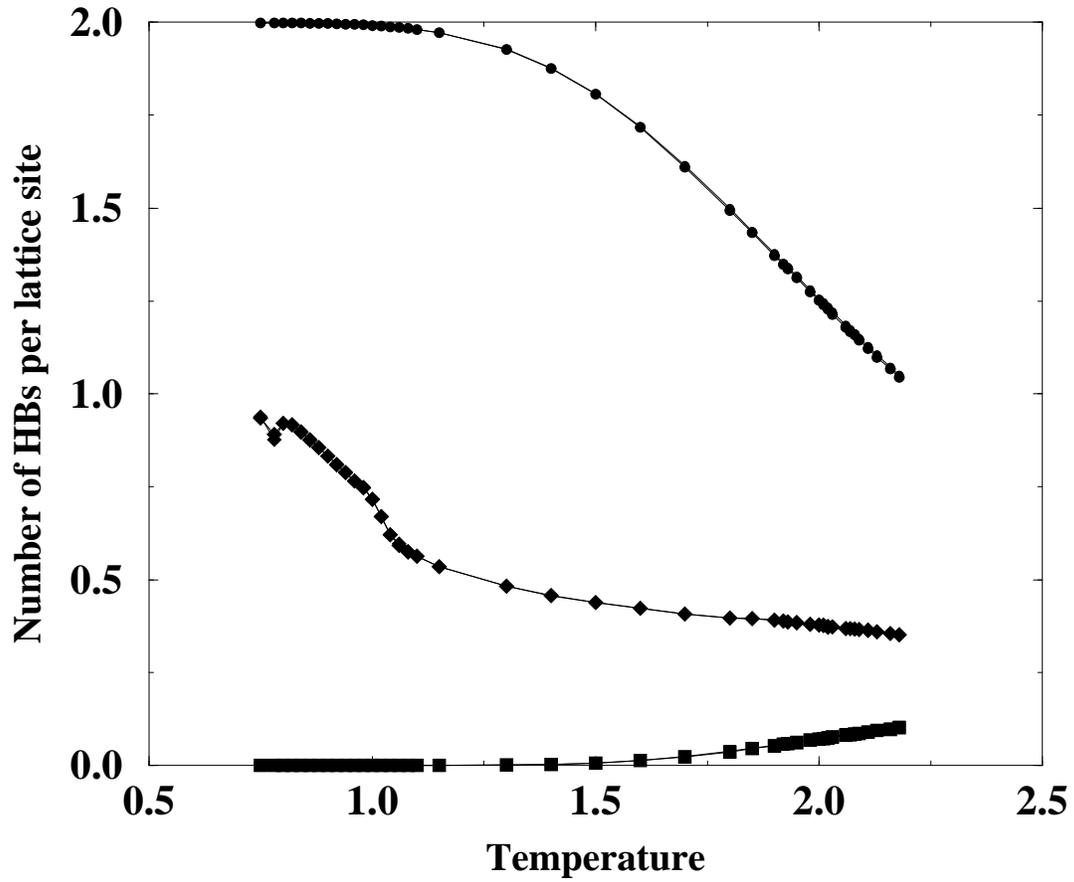,width=16cm,angle=-90}}

\vspace{0.2in}
\caption{Number of hydrogen bonds per site in the $D=16$ film with $k_w=0.0134e^{10/T^{\star}}$ 
vs. $T^{\star}$.
Circles-- bonds between the walls and the surface layers ($z=0$, and $z=15$), 
squares--bonds within the surface layers, 
and diamonds-- in--layer bonds in the middle layers ($z=7$, and $z=8$).} 
\label{HBtotals}
\end{figure}

A way to qualitatively understand these findings
is as follows. When the walls are non--interacting, the 
only effect of confinement is the loss of neighbors for molecules in the
surface layer.
Since this reduces the net
unfavorable interaction energy of the system it stabilizes
the single phase. In contrast, in the
case of interacting walls, HB interactions occur between the
wall and the molecules. Since each molecule has only 
one HB donor and one HB acceptor,
the number of HBs between molecules  within the surface layer is
reduced due to the presence
of the interacting walls. This effect is shown
in Figure \ref{HBtotals}, where we plot the average number of 
molecule-molecule and molecule-wall HB contacts for the surface layer
where $k_w=0.0134e^{10/T^{\star}}$.
Since molecule-molecule HBs are one of the strong factors 
aiding the miscibility of these systems, and are
primarily responsible for the LCST, it is clear that the single phase
must be destabilized in the thin films. 
In fact, at the lowest temperatures in Figure \ref{HBtotals}
all donors and acceptors on molecules in the 
surface layer are occupied
by the HB's with the walls.
Thus, only isotropic nearest neighbor
interactions occur  between these molecules, as well as with molecules
in the adjacent layer (i.e., $z=1$).
Since  a bulk system with isotropic nearest
neighbor interactions only possesses a UCST, to a zeroth
approximation, the surface will remain ordered at all low
temperature conditions. 

In summary, we have presented results of 
Monte Carlo simulations which show that the phase behavior of
an associated fluid mixture in a thin film geometry can be 
qualitatively different from simple mixtures, which show
only ``ordinary" behavior. The phase behavior of such HB
mixtures, which are defined by the balance of specific and dispersive
interactions, can be disturbed at interfaces, 
leading to the occurrence of a surface
transition.  
This finding strikingly rationalizes experimental
results where surface transitions appear to be ubiquitous.

The financial support from the American Chemical Society [ACS-PRF]
and the National Science Foundation [CTS-9311915 and CTS-9704907]
are gratefully acknowledged. We thank Boris Veytsman
for many long and useful discussions.


\begin{thebibliography}{10}

\bibitem{emailmike}
e-mail address:kotelyan@plmsc.psu.edu.

\bibitem{emailkumar}
e-mail address:kumar@plmsc.psu.edu.

\bibitem{Fish0}
H.~Nakanishi and M.E. Fisher.
\newblock {\em Phys. Rev. Lett.}, 49:1565, 1982.

\bibitem{Fish1}
M.E. Fisher.
\newblock In M.S.Green, editor, {\em Critical Phenomena}, volume~51 of {\em
  Proceedings of the Enrico Fermi International School of Physics}, New York,
  1971. Academics.

\bibitem{Fish2}
M.E. Fisher and H.~Nakanishi.
\newblock {\em J. Chem. Phys.}, 75:5857, 1981.

\bibitem{NakanishiFisher}
H.~Nakanishi and M.~Fisher.
\newblock {\em J. Chem. Phys.}, 78:3279, 1983.

\bibitem{Fish3}
H.~Nakanishi and M.E. Fisher.
\newblock {\em J. Phys. C.}, 16:L95, 1983.

\bibitem{Mader76}
S.~Mader.
\newblock {\em Thin Solid Films}, 35:195, 1976.

\bibitem{Meadows}
M.~R. Meadows, B.~A. Scheibner, R.C. Mockler, and W.J. O'Sullivan.
\newblock {\em Phys. Rev. Lett.}, 43:592, 1979.

\bibitem{Scheibner}
B.~A. Scheibner, M.~R. Meadows, R.C. Mockler, and W.J. O'Sullivan.
\newblock {\em Phys. Rev. Lett.}, 43:590, 1979.

\bibitem{PSPVME}
S.~Reich and Y.~Cohen.
\newblock {\em J. Polym. Sci. Poly. Phys. Ed.}, 19:1255, 1981.

\bibitem{russians}
V.M. Andreyeva, A.I. Dolinnyy, A.B. Burdin, and Kozhevnikova Y.A.
\newblock {\em Flu. Mech. Res.}, 21:72, 1992.

\bibitem{BinderReview}
K.~Binder.
\newblock {\em Phase Transitions}, 8:2, 1983.

\bibitem{Lubensky}
T.C. Lubensky and M.H. Rubin.
\newblock {\em Phys. Rev. B}, 12:3885, 1975.

\bibitem{Hirshfelder}
J.~Hirschfelder, D.~Stevensen, and H.~Eyring.
\newblock {\em J.Chem.Phys.}, 5:896, 1937.

\bibitem{Barker52}
J.A. Barker.
\newblock {\em J.Chem.Phys.}, 20:1526, 1952.

\bibitem{wheeler}
J.~Wheeler.
\newblock {\em J. Chem. Phys.}, 62:433, 1975.

\bibitem{walker}
J.S. Walker and C.A. Vause.
\newblock {\em J.Chem.Phys.}, 79:2660, 1983.

\bibitem{Boris89}
B.A. Veytsman.
\newblock {\em J.Phys.Chem.}, 94:8499, 1990.

\bibitem{desai}
C.~Sagui, A.M. Somoza, C.~Roland, and R.C. Desai.
\newblock {\em J. Phys. A. Math. Gen.}, 26:L1163, 1993.

\bibitem{kardar}
H.~Li, M.~Paczuski, M.~Kardar, and K.~Huang.
\newblock {\em Phys. Rev. B.}, 44:8274, 1991.

\bibitem{entropy}
In this model we have assumed that the molecules are unstructured, and
  consequently the donor and acceptor sites can be located at arbitrary points.
  Since the HB formation leads to a reduction in the entropy of the molecules,
  a factor which is not taken into account $a-priori$, we have introduced a
  prefactor that accounts for this factor in the definition of the equilibrium
  constants.

\bibitem{MKBulk}
M.~Kotelyanskii, B.~Veytsman, and S.K. Kumar.
\newblock {\em in preparation}, 1997.

\bibitem{SemiGrand}
A.~Sariban and K.~Binder.
\newblock {\em J.Chem.Phys.}, 86:813, 1987.

\bibitem{BinderHeermann}
K.~Binder and D.W. Heermann.
\newblock Springer--Verlag, Berlin Heidelberg New York London Paris Tokyo,
  1988.

\end{thebibliography}

\clearpage

\newpage

\newpage

\newpage

\end{document}